\documentclass[12pt,english,aps,prb,preprint,showpacs]{revtex4}
\usepackage{amsmath}
\usepackage{graphicx}
\providecommand{\tabularnewline}{\\}
\usepackage{bm}
\usepackage{babel}
\makeatother

\begin{document}

\title{The effect of attractions on the local structure of liquids and colloidal fluids}

\author{Jade Taffs}

\affiliation{School of Chemistry, University of Bristol, Bristol, BS8
1TS, UK}

\author{Alex Malins}

\affiliation{School of Chemistry, University of Bristol, Bristol, BS8
1TS, UK}

\affiliation{Bristol Centre for Complexity Science, University of Bristol, Bristol, BS8
1TS, UK}

\author{Stephen R. Williams}

\affiliation{Research School of Chemistry, The Australian National University,
Canberra, ACT 0200, Australia}

\author{C. Patrick Royall}

\affiliation{School of Chemistry, University of Bristol, Bristol, BS8
1TS, UK}

\affiliation{paddy.royall@bristol.ac.uk}

\email{paddy.royall@bristol.ac.uk}
\date{\today}

\begin{abstract}
We revisit the role of attractions in liquids and apply these concepts to colloidal suspensions. Two means are used to 
investigate the structure; the pair correlation function and a recently developed topological method.
The latter identifies structures topologically equivalent to ground state clusters formed by isolated groups of $5\leq m \leq 13$
particles, which are specific to the system under consideration. Our topological methodology shows
that, in the case of Lennard-Jones, the addition of attractions increases the system's ability to form larger ($m\geq8$) clusters, although pair-correlation functions are almost identical. Conversely, in the case of short-ranged attractions,
pair correlation functions show a significant response to adding attraction, while the liquid structure
exhibits a strong decrease in clustering upon adding attractions.
Finally, a compressed, weakly interacting system shows a similar pair structure and topology.
\end{abstract}


\maketitle

\section{Introduction}

\label{sec:Introduction}

Among the cornerstones of our understanding of the structure of bulk simple liquids
is that it is dominated by the repulsive core. This leads to the idea that hard spheres
form a suitable basic model of the liquid state. 
The liquid pair structure may then be accurately calculated using, for example
the Percus-Yevick closure to the Ornstein-Zernike equation for hard spheres,
and treating the remainder of the interaction as a perturbation \cite{barker1976,weeks1971,chandler1983}.

Although in principle colloidal dispersions are rather complex multicomponent
systems, the spatial and dynamic asymmetry between the colloidal particles
(10 nm-1 $\mu$m) and smaller molecular and ionic species has enabled
the development of schemes where the smaller components are formally integrated out
\cite{likos2001}. This leads to a one-component picture, where only
the \emph{effective} colloid-colloid interactions need be considered.
The behaviour in the original complex system may then be faithfully
reproduced by appealing to liquid state theory~\cite{hansen} and
computer simulation \cite{frenkel}. Since the shape of the particles
is typically spherical, and the effective colloid-colloid interactions
may be tuned, it is often possible to use models of simple liquids
to accurately describe colloidal dispersions. 

In colloidal systems, due to the mesoscopic length- and longer time-scales, 
one may also determine the structure in real space 
in 2D and 3D at the single
particle level using optical microscopy 
\cite{royall2003,brunner2002} and optical tweezers \cite{crocker1994}. This may be done with sufficient precision
that interaction potentials can be accurately determined both for
purely repulsive systems \cite{brunner2002,royall2006} and for systems
with attractive interactions \cite{royall2007}.


\begin{figure}
\begin{centering}
\includegraphics[width=9cm]{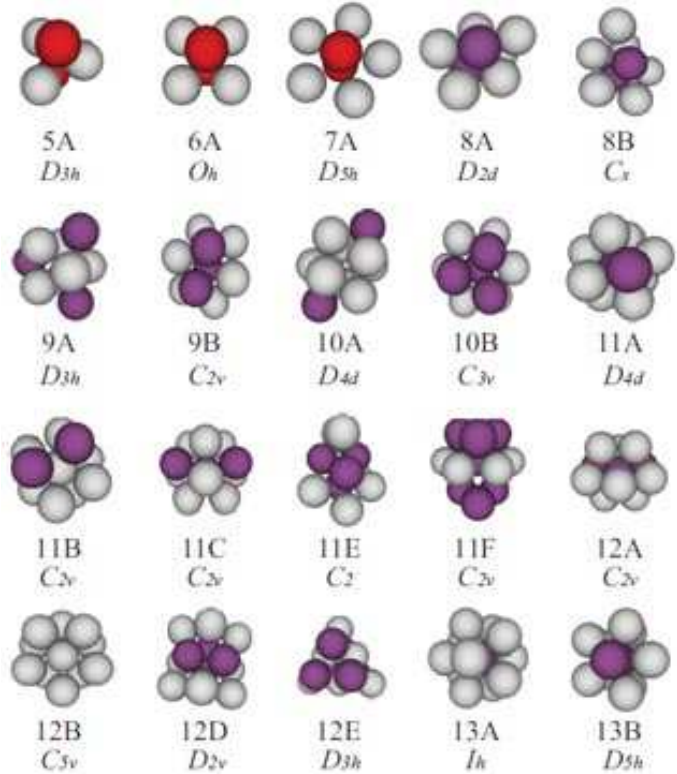} 
\par\end{centering}
\caption{(color online) Clusters found in bulk systems using the topological
cluster classification. \textbf{For $m\leq7$, where $m$ is the number of particles in a cluster,} all studied ranges of the Morse potential equation (\ref{eqMorse}) form
clusters of identical topology. In the case of larger $m$ the cluster
topology depends on the interaction range. Here we follow the nomenclature
of Doye \emph{et. al.} \cite{doye1995} where $A$ corresponds to
long-ranged potentials and $B.$... to minimum energy clusters of shorter-ranged potentials.}
\label{figTCC} 
\end{figure}

\begin{figure}
\begin{centering}
\includegraphics[width=14.0cm]{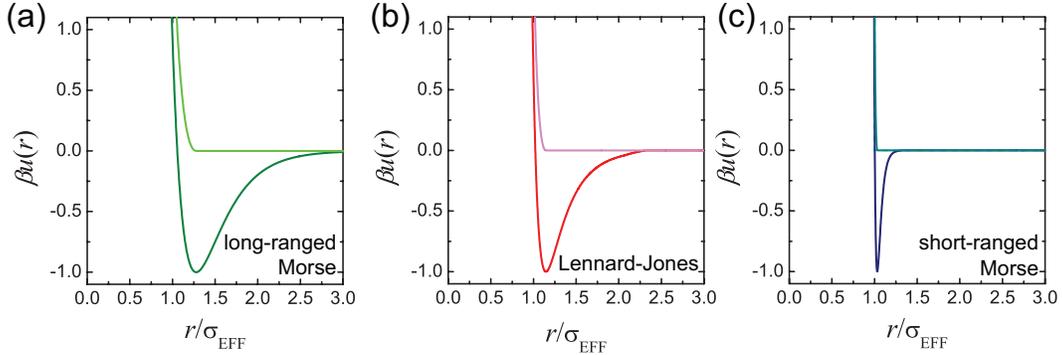} 
\par\end{centering}
\caption{(color online) Interaction potentials used. (a)
Long-ranged potentials: Morse 
(dark green) and truncated Morse (bright green) 
with range parameter $\rho_0=4.0$.
(b) Lennard-Jones (red) and WCA (pink). 
(c) Short-ranged potentials: Morse
(blue) and truncated Morse (turquoise) with range parameter $\rho_0=25.0$. 
Dashed cyan line in (c) denotes the hard
sphere interaction. $\sigma_{EFF}$ denotes the effective hard sphere
diameter as defined in equation (\ref{eqEffective}) and listed in table \ref{tableStatePoints}.}
\label{figU}
\end{figure}

It has been conjectured as far back as the 1950s that the structures
formed by clusters of small groups of particles in isolation might
be prevalent in liquids \cite{frank1952, coslovich2007}. 
More recently it has been
demonstrated that 
for spherically symmetric
attractive interactions,
the structure of clusters of size $m>7$ particles
depends upon the range of the potential,  as shown in Fig. \ref{figTCC} \cite{doye1995}. 
This brings a natural question: if the structures 
defined by these clusters
are indeed prevalent in liquids, and they depend upon the range of
the interaction, then might liquids with differing interaction ranges
exhibit differing cluster populations? Moreover, while
removing the attractive component of Lennard-Jones [Fig. \ref{figU}(b)]
has little effect on the pair structure \cite{barker1976,weeks1971}, what is the effect on any cluster
population?

We have recently developed a novel means to identify structure in
simple liquids. In isolation, small groups of particles 
form clusters with well-defined
topologies. These have been identified for the Lennard-Jones potential
\cite{wales1997} and for the Morse potential, which has a variable
range \cite{doye1995}. We identify clusters
relevant to the Lennard-Jones and Morse Potentials in bulk liquids,
with a method we term the Topological Cluster Classification (TCC)
\cite{williams2007}. Here we use this scheme as a highly sensitive
probe of the liquid structure. It is our intention to use the
TCC to investigate possible differences in structure between the Lennard-Jones liquid
and that resulting from the repulsive part of the Lennard-Jones interaction, 
the Weeks-Chandler-Andersen (WCA) potential \cite{weeks1971}.
Although we have argued that some colloidal liquids are
well described by a short-ranged Morse potential \cite{royall2008gel}, the 
structure of clusters of adhesive hard spheres has very recently been shown to
exhibit some degeneracy, with multiple cluster topologies having the same number of
bonds in the limit of short-ranged attractions \cite{arkus2009, meng2010}. However, minimising the 
second moment (or radius of gyration) of clusters of hard spheres \cite{sloane1995,manoharan2003} shows a strong
similarity with the short-ranged Morse system \cite{doye1995}.
Our TCC methodology has some similarities to the common neighbour analysis
introduced by Andersen \cite{jonsson1988, honeycutt1987}, however here we 
focus on clusters rather than bonds.

Since the tunability
of colloidal systems allows a wide range of potentials to be realised, including long-ranged
interactions relevant to metals \cite{wette2009, royall2006, doye1995, yethiraj2003},
we also consider long-ranged (Morse) potentials, in addition to the Lennard-Jones
interaction and short-ranged Morse potential, along with their purely repulsive counterparts. We further compare
with hard spheres. In these systems, we study the 
groups of particles topologically equivalent to
ground state clusters found in isolation.

The canonical model of colloid-polymer mixtures of Asakura and Oosawa, assumes hard sphere 
colloid-colloid and colloid-polymer interactions, while the polymer-polymer interaction is ideal \cite{asakura1954}. 
A one-component description \cite{asakura1958,dijkstra1999}, accurate for small polymer-colloid size ratios 
\cite{dijkstra1999} leads to a hard core with a short-range attraction. We have recently shown that, for the
parameters we shall consider here, the continuous Morse potential provides a reasonably accurate description
of this system \cite{taffs2010}. Meanwhile, longer interaction ranges correspond to metals \cite{doye1995} 
and purely repulsive long-ranged interactions are relevant to soft matter systems such as
charged colloids, star polymers \cite{likos2006}, star polyelectrolytes \cite{jusufi2009} and
colloidal microgel particles. 

This paper is organised as follows. In section \ref{sec:Simulations-and-Interaction}
we describe the simulation methodology and our approach for comparing
different interaction potentials, our results are presented in section
\ref{sec:Results} and we conclude with a discussion in section \ref{sec:Conclusions}.
Our main results can be summarised as follows. Although Lennard-Jones shows very little
change in the radial distribution function $g(r)$ upon adding attractions, the topology is significantly altered: 
adding attractions promotes the formation of larger clusters. Conversely, short-ranged systems 
show the opposite behaviour: adding attractions strongly decreases clustering, while the first peak of $g(r)$ shows some increase upon adding attractions.

\section{Simulations and Interaction Potentials}

\label{sec:Simulations-and-Interaction}

We use standard Monte-Carlo (MC) simulation in the NVT ensemble \cite{frenkel} with 
$N=2048$ particles. Each simulation run is equilibrated for
$2\times10^{7}$ MC moves and run for up to a further $10^{8}$ moves.
In all cases, we confirmed that the system was in equilibrium on the
simulation timescale by monitoring the potential energy. 

\subsection{Interaction potentials}
\label{interactionPotentials}

We seek to compare systems with different interactions, under similar
conditions. Weeks, Chandler and Andersen \cite{weeks1971} provided a protocol by which the
Lennard-Jones potential could be compared with the so-called WCA potential (Lennard-Jones
without attractions). The pair interaction $u(r)$ is separated into
two parts: \[u(r)=u_{0}(r)+w(r)\] where $r$ is the separation between particles, $u_{0}(r)$ is the reference (repulsive) interaction and $w(r)$
is the perturbative attraction. In the Lennard-Jones case, 

\begin{equation}
\beta u_{LJ}(r)=4\beta \varepsilon_{LJ}\left[\left(\frac{\sigma}{r}\right)^{12}-\left(\frac{\sigma}{r}\right)^{6}\right]
\label{eqLennardJones}
\end{equation}

\noindent where $\beta=1/k_{B}T$ where $k_{B}$
is Boltzmann's constant and $T$ is temperature. Here $\varepsilon_{LJ}=1/T$
is the well depth. WCA thus define the reference potential as 

\begin{equation} 
\beta u_{WCA}(r)=\begin{cases}
\text{$4\beta \varepsilon_{LJ}\left[\left(\frac{\sigma}{r}\right)^{12}-
\left(\frac{\sigma}{r}\right)^{6}\right] + \beta \varepsilon_{LJ}$} & \text{for $r \le 2^{1/6} \sigma,$}\\ 
\text{0}&\text{for $r > 2^{1/6} \sigma.$}\\
\end{cases}
\label{eqWCA}
\end{equation}

\noindent 
2nd order perturbation theories \cite{barker1976} allow accurate prediction of the pair structure. However,
here we are interested in a particle based analysis that probes the structure at a level beyond the two body distribution function and restrict ourselves to the interactions given by equations 
(\ref{eqLennardJones}) and (\ref{eqWCA}).

In the case of the longer and shorter ranged interactions, we use the Morse potential which reads

\begin{equation}
\beta u_{M}(r)=\beta \varepsilon_{M}e^{\rho_{0}(\sigma-r)}(e^{\rho_{0}(\sigma-r)}-2)
\label{eqMorse}
\end{equation}

\noindent where $\rho_{0}$ is a range parameter and $\beta\varepsilon_{M}$
is the potential well depth. 
We set $\rho_0=25.0$ to simulate a system with short-ranged attractions similar to a colloid-polymer mixture
and $\rho_0=4.0$ as an example of a longer-ranged system.
Following the WCA approach,
we introduce a repulsive (truncated) Morse potential, which is also truncated at the minimum and is defined as
follows. The potentials we use are plotted in Fig. \ref{figU}.

\begin{equation} 
\beta u_{TM}(r)=\begin{cases}
\text{$\beta\varepsilon_{M}e^{\rho_{0}(\sigma-r)}(e^{\rho_{0}(\sigma-r)}-2)+
\beta\varepsilon_{M} $}&\text{ for $r \le \sigma,$}\\ 
\text{0}&\text{for $r > \sigma,$}\\
\end{cases} \label{eqTMR}
\end{equation}

\noindent This truncated Morse potential is thus similar to hard spheres for $\rho_0=25.0$ [Fig. \ref{figU}(c)].
The repulsive systems have well-defined truncations, $2^{1/6}\sigma$ for WCA and $\sigma$ for the Morse potential.
In the case of the attractive systems, we truncate and shift both
Lennard-Jones and Morse ($\rho_0 = 25.0$) at $2.5 \sigma$ and the long-ranged Morse ($\rho_0 = 4.0$) at
$4.0 \sigma$.

\subsection{Comparing different systems}

We have outlined a means by which we can compare
systems with and without attraction, by removing the attractive
part of the interaction. In order to match state points
between systems with differing interaction ranges, we use the extended
law of corresponding states introduced by Noro and Frenkel \cite{noro2000}.
Specifically, this requires two systems to have identical reduced
second virial coefficients $B_{2}^{*}$ where 

\begin{equation}
B_{2}^{*}=\frac{B_{2}}{\frac{2}{3}\pi\sigma_{EFF}^{3}}
\label{eqReducedVirial}
\end{equation}

\noindent where $\sigma_{EFF}$ is the effective hard sphere diameter and the
second virial coefficient

\begin{equation}
B_{2}= 2 \pi \intop^{\infty}_0 drr^{2}\left[1-exp\left(-\beta u(r)\right)\right].
\label{eqCorrespondingStates}
\end{equation}

The effective hard sphere diameter is defined as

\begin{equation}
\sigma_{EFF}=\intop^{\infty}_0 dr\left[1-exp\left(-\beta u_{REP}(r)\right)\right]
\label{eqEffective}
\end{equation}

\noindent where the repulsive part of the potential
$u_{REP}$ is described above in section \ref{interactionPotentials}. 
Thus we compare different interactions by equating $B_{2}^{*}$ and
$\sigma_{EFF}$. The latter condition leads to a constraint on number density
\begin{equation}
\rho_{EFF}=\frac{N\pi\sigma_{EFF}^{3}}{6V}
\label{eqDensity}
\end{equation}

\noindent where $V$ is the volume of the simulation box. We fix $\rho_{EFF}$
to a value equivalent to the Lennard-Jones triple point ($\rho_{LJ}=0.85$) throughout.
In the case of hard spheres, this value is $\rho_{HS}\approx0.9310$ or $\phi_{HS}=\pi\rho_{HS}/6 \approx0.4875$ where
$\phi$ is the packing fraction.
Details of state points investigated are given in table \ref{tableStatePoints}.



\subsection{The Topological Cluster Classification}

To analyse the structure, we first identify the bond network using a modified Voronoi
construction with a maximum bond length $r_{c}=1.8\sigma$ and four-membered ring 
parameter $f_{c}=0.82$ (TCC) paper\cite{williams2007}. Having identified the bond network, 
we use the TCC to determine the nature of the cluster.
This analysis identifies all the shortest path three, four and five
membered rings in the bond network. We use the TCC to find clusters
which are global energy minima of the Lennard-Jones and Morse potentials.
We identify all topologically distinct Morse clusters,
of which the Lennard-Jones clusters form a subset (the Lennard-Jones and Morse
interactions are similar in the case that the range
parameter $\rho_{0}=6.0$ and the topology of the ground state clusters is identical). In addition
we identify the FCC and HCP thirteen particle structures in terms
of a central particle and its twelve nearest neighbours. We illustrate
these clusters in Fig. \ref{figTCC}. 
In the case of the Morse potential, for $m>7$ there is more than
one cluster which forms the ground state, depending on the range of
the interaction \cite{doye1995}. 
We therefore consider ground state clusters for each
system and, separately, calculate all topologically distinct
Morse clusters for $m<14$.
For more details see \cite{williams2007}.

To compare the various fluids we study here, we proceed as follows.
Comparing systems with and without attractions, we consider the ground state 
clusters (of the attractive system). If a particle is a member of more than one cluster,
it is taken to `belong' to the larger cluster. Thus, the total cluster
population $\leq N$ the total number of particles. However, when we seek
to compare different potentials, we need to account for the fact that these may have
different ground state clusters. If the particle is part of
two clusters which are different in size, we choose to count it as the larger cluster, but if
the particle is part of two clusters of the same size, it is counted as the cluster corresponding to the
shorter-ranged interaction. In this case, the total number of particles counted as belonging to a cluster
can exceed the number of particles in the simulation.

\subsection{Systems studied}

The different systems considered are listed in Table \ref{tableStatePoints}. In addition
to the state point ($\varepsilon, \rho$), we list the reduced second virial coefficient $B^*_2$ 
and effective hard sphere diameter $\sigma_{EFF}$.
Some comments upon the use of clusters in the case of repulsive
systems are in order. Clearly, isolated clusters require cohesive
forces, however here we compare the WCA and the truncated and shifted
Morse potential with their cohesive counterparts
and we assume it is appropriate to consider the same clusters. Given the similarity
of the truncated Morse potential to hard spheres [Fig.
\ref{figU}(c)], it is instructive to include these also. 

We are motivated to consider the Lennard-Jones
triple point, as we expect clusters to be more prevalent at lower temperature. 
However, mapping the short-ranged Morse potential ($\rho_0=25.0$) to the Lennard-Jones triple point 
leads to a system unstable to crystallisation. We found that at higher temperature the Morse system
was stable (on the timescales of these simulations) against crystallisation
for $\beta\varepsilon_{M}=2.0$. Thus we compare high-temperature Lennard-Jones (T=2.284)
with Morse $\rho_0=25.0$ and triple point Lennard-Jones with the long-ranged Morse $\rho_0=4.0$.

\begin{table}
\caption{State points studied. LJ high temp. and triple correspond to the two temperatures
at which Lennard-Jones and WCA simulations were carried out.
Trunc. Morse denotes the truncated Morse interaction (equation \ref{eqTMR}).}
\begin{tabular}{cccccc}
\hline 
System&
$B_{2}^{*}$&
$\beta\varepsilon$&
$\rho$ &
$\sigma_{EFF}$ \tabularnewline
\hline
\hline 
LJ high temp.&  			
-0.2325&
0.4447&
0.9776&
0.9839  \tabularnewline
\hline 
WCA high temp.&  		
2.013&
0.4447&
0.9776&
0.9839 \tabularnewline
\hline 
Morse $\rho_0 = 25.0$ &  		
-0.2325&
2.0&
0.9837 &
0.9818 \tabularnewline
\hline 
Trunc. Morse  $\rho_0 = 25.0$&   
0.9818&
2.0&
0.9837 &
0.9818 \tabularnewline
\hline 
Hard Spheres&
1.0&
N/A&
0.9310 &
1.0 \tabularnewline
\hline
LJ triple&  			
-3.742 &
1.471 &
0.85 &
1.0308 \tabularnewline
\hline
WCA triple&  
2.307 &   			
1.471 &
0.85 &
1.0308  \tabularnewline
\hline
Morse $\rho_0 = 4.0$ &  
-3.742 &
0.9109 &
1.548 &
0.8441 \tabularnewline
\hline
Truncated Morse $\rho_0 = 4.0$ &  
1.015  &
0.9109 &
1.548 &
0.8441 \tabularnewline
\hline
\end{tabular}
\label{tableStatePoints}
\end{table}

\section{Results and Discussion}

\begin{figure}
\begin{centering}
\includegraphics[width=14cm]{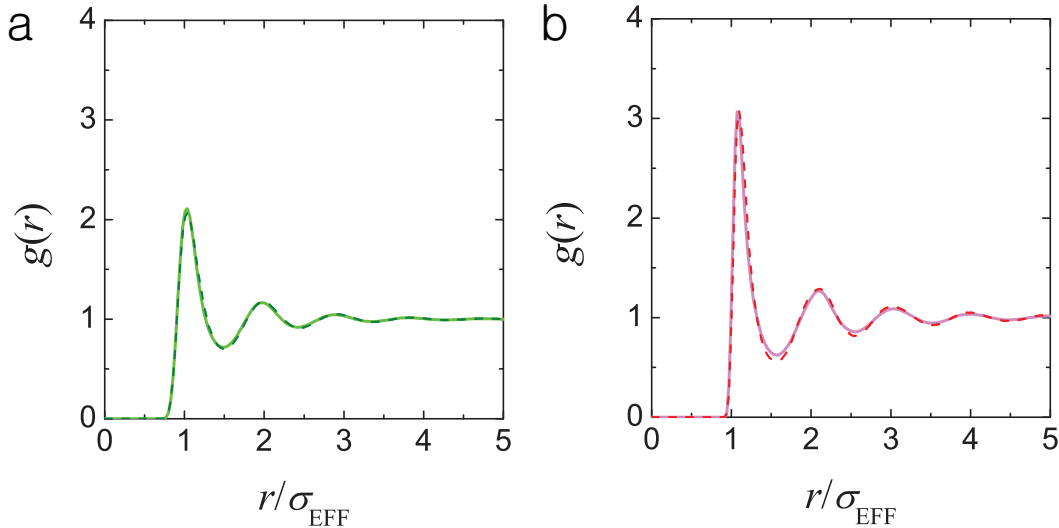} 
\par\end{centering}
\caption{(color online) Pair-correlation functions. (a) Long-ranged potentials:
Morse $\rho_0=4.0$ with (dark green, dashed) and without (bright green) attractions. Here  
$\beta\varepsilon_{M} = 0.9109$.
(b) Lennard-Jones (red, dashed) and WCA (pink) for a well depth of $\beta\varepsilon_{LJ}=1.471$ (the triple point).}
\label{figGTriple} 
\end{figure}

\label{sec:Results}





\subsection{The Lennard-Jones Triple Point: Long-ranged interactions}

We take as our starting point for the analysis of these data the result that for dense liquids, the WCA potential readily captures the pair structure of the Lennard-Jones liquid \cite{weeks1971}. The radial distribution functions $g(r)$ for the Lennard-Jones liquid at the triple point and the corresponding WCA system are plotted in Figure \ref{figGTriple}(b). The effectiveness of WCA in describing the pair structure is clear. The same observation holds for the longer-ranged ($\rho_0=4.0$) Morse and truncated Morse systems shown in Figure \ref{figGTriple}(a). 

\begin{figure}
\begin{centering}
\includegraphics[width=15cm]{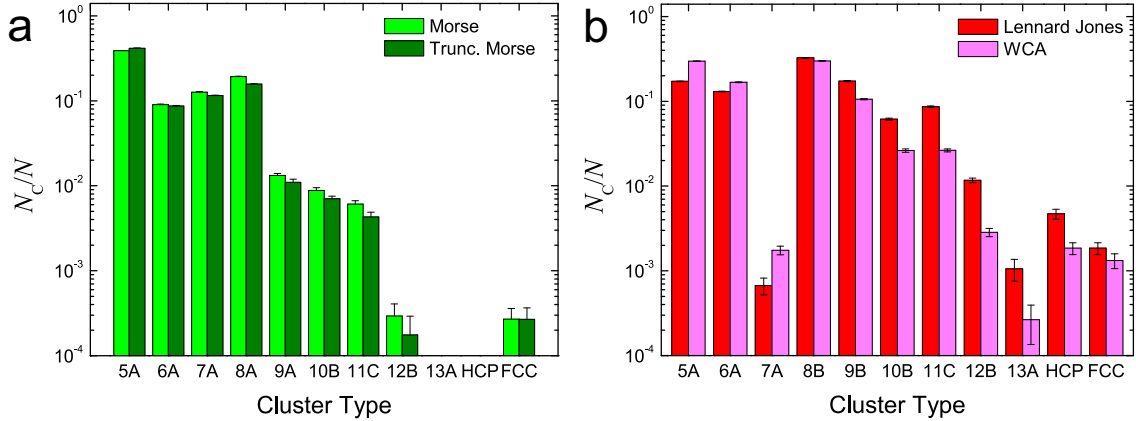} 
\par\end{centering}
\caption{(color online) Population of particles in a given cluster. $N_c$ is the number of particles in a given cluster,
$N$ the total number of particles sampled. Here we consider only ground state clusters for each system. (a) Morse ($\rho_0=4.0$) (dark green)
truncated Morse (bright green). (b)
Lennard-Jones at the triple point (red) and corresponding WCA (pink). Note the semi-log scale.}
\label{figClusGroundLong} 
\end{figure}

\begin{figure}
\begin{centering}
\includegraphics[width=15.0cm]{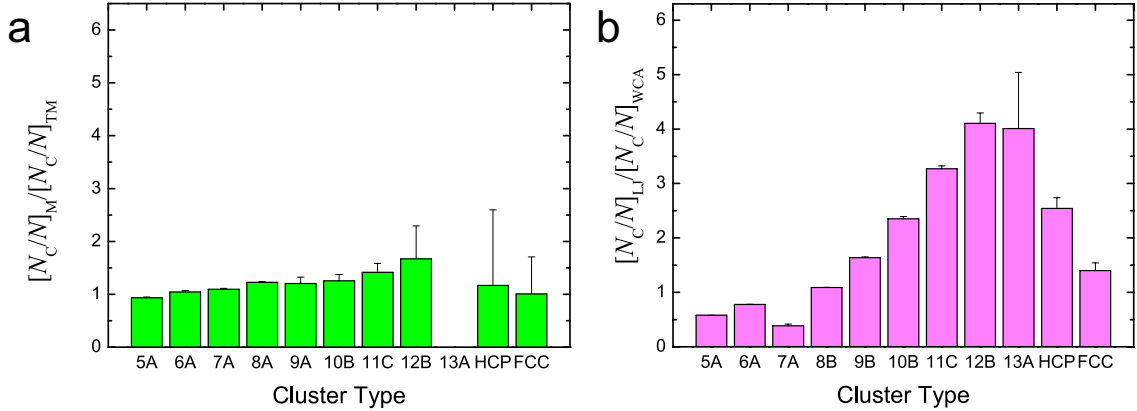} 
\par\end{centering}
\caption{(color online) Ratio of cluster populations in systems mapped to the Lennard-Jones triple point. (a) Morse and truncated Morse ($\rho_0=4.0$) and (b) Lennard-Jones and WCA. These plot the same data as Fig. \ref{figClusGroundLong} expressed to emphasize the difference between the systems.}
\label{figRatio} 
\end{figure}

Turning to the TCC analysis, in the WCA-Lennard-Jones system we see a somewhat different story, as shown in Fig. \ref{figClusGroundLong}(b).
Note the logarithmic scale in this plot: cluster populations vary over three orders of magnitude, clear relative differences
are seen between Lennard-Jones and WCA.
Unlike the $g(r)$ which are very similar, there is a clear trend in the cluster populations $N_c/N$. Larger clusters are more prevalent in the Lennard-Jones system, compared to the WCA. The differences are emphasized in Fig. \ref{figRatio}(b) which plots the ratio of the cluster populations in Fig. \ref{figClusGroundLong}(b). One might argue that smaller clusters may readily be formed simply by compressing spheres. 
The addition of attractions promotes the formation of larger clusters which require more organisation and co-operativity. We remark that the difference in structure revealed by the TCC is rather significant, given that the radial distribution functions are so similar. For example, there is a twofold difference in the  triangular bipyramid 5A, one of the most popular clusters. 
As for the 13A icosahedron, its population is quadrupled by adding attractions.
Note also that in these equilibrium liquids, we find a small but measurable number of particles with local crystalline topology, even though there is no sign of splitting in the second peak of $g(r)$ (Fig. \ref{figGTriple}), which is often taken to be a sign that the liquid is close to crystallising \cite{truskett1998}. Of the smaller clusters, the 7A pentagonal bipyramid is found in limited quantities. However, it is found also as part of all the larger clusters which form minima for the Lennard-Jones except HCP and FCC so our counting methodology counts some 7A ($D_{5h}$) particles as members of larger clusters.

The longer-ranged Morse ($\rho_0=4.0$) system on the other hand shows very little difference upon adding attractions.
Due to its softness, the long-ranged Morse system has a rather small value of $\sigma_{EFF}$ (table \ref{tableStatePoints}). Thus, matching Lennard-Jones requires a higher density, which leads to some overlap of the particles. 
For example the mean inter-particle spacing $d_m=\rho^{-1/3}\approx 0.8644\sigma$. The Morse potential has its minimum located at $\sigma$. However, such is the long ranged nature of Morse $\rho_0=4.0$ that in fact, $0.8644\sigma$ remains within the attractive well, although there is some compression. Thus both with and without attractions, the system is compressed, which may dominate the local structure.
Furthermore, the value of $\beta \varepsilon_M=0.9109$ (table \ref{tableStatePoints}) indicates that the interactions here are rather weak. As we shall see below, weaker interactions can lead to topologically similar structures. Note the relatively high abundance of the pentagonal bipyramid 7A, due to fewer reclassifications as higher-order clusters. Plotting the ratio of the cluster populations [Fig. \ref{figRatio}(a)] further emphasizes the similarity of the long-ranged Morse interaction with and without attractions: the ratio lies close to unity in all cases, except the 13A icosahedron where the statistics are insufficient for a reliable comparison.

\begin{figure}
\begin{centering}
\includegraphics[width=15cm]{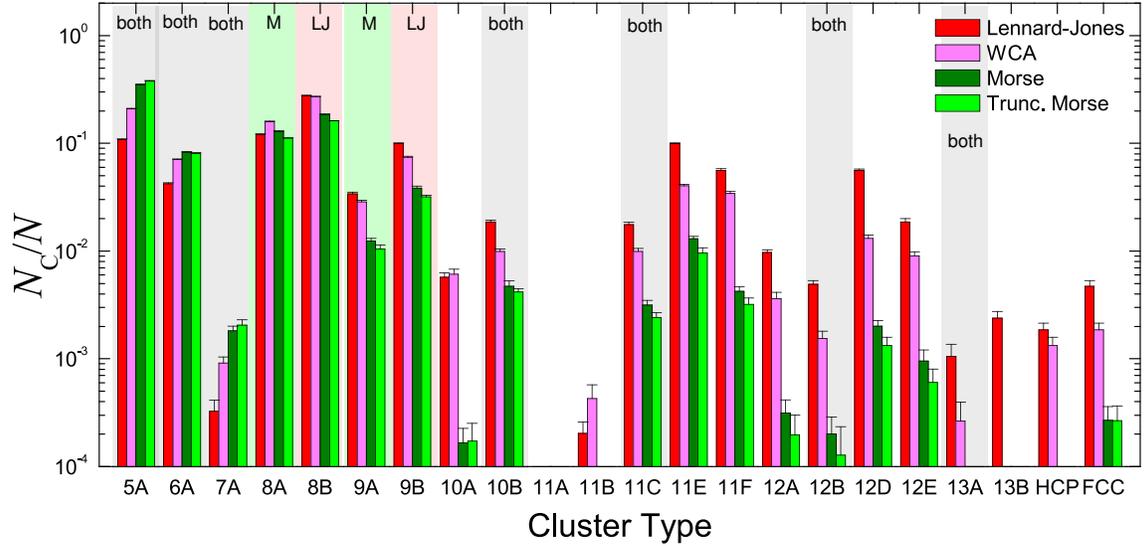} 
\par\end{centering}
\caption{(color online) Population of particles in a given cluster at parameters mapped to the Lennard-Jones triple point. $N_c$ is the number of particles in a given cluster,
$N$ the total number of particles sampled. Here we consider ground state clusters for all ranges of the Morse potential \cite{doye1995}. Colours are Lennard-Jones (red), corresponding WCA (pink), Morse ($\rho_0=4.0$) (bright green) and truncated Morse (dark green).
Those clusters which are ground states are labelled as `both' when both potentials share the same ground state, and
`LJ'  and `M' corresponding to the Lennard-Jones and Morse cases accordingly. Note the semi-log scale.}
\label{figAllLongColdSort} 
\end{figure}

We now plot the population of all identified clusters. This enables us to directly compare the populations of the long-ranged Morse and Lennard-Jones systems. 
We see that there is no strong preference for ground states [shaded in Fig. 
(\ref{figAllLongColdSort})]. In fact a number of ground states are less populated than other clusters of the same size.
Comparing the different systems, the general trend is of LJ/WCA tending to form larger clusters than the long-ranged Morse, which is consistent with the idea that the long-ranged Morse is a weakly interacting, compressed fluid. 
Recall that, for example in the case of 11-membered clusters, we count a given particle as a member of an 11F ($C_{2v}$) if it is 
a member of more than one $m=11$ clusters of which one is an 11F. While this may inflate the populations of such clusters, we argue that systems with differing ground states are compared in an unbiased way. One result of considering all Morse clusters is that Lennard-Jones has by far the largest number of 13A icosahedra, although the population of the 13B decahedral cluster is larger.

\subsection{High-Temperature systems : Short-ranged interactions}

\begin{figure}
\begin{centering}
\includegraphics[width=14cm]{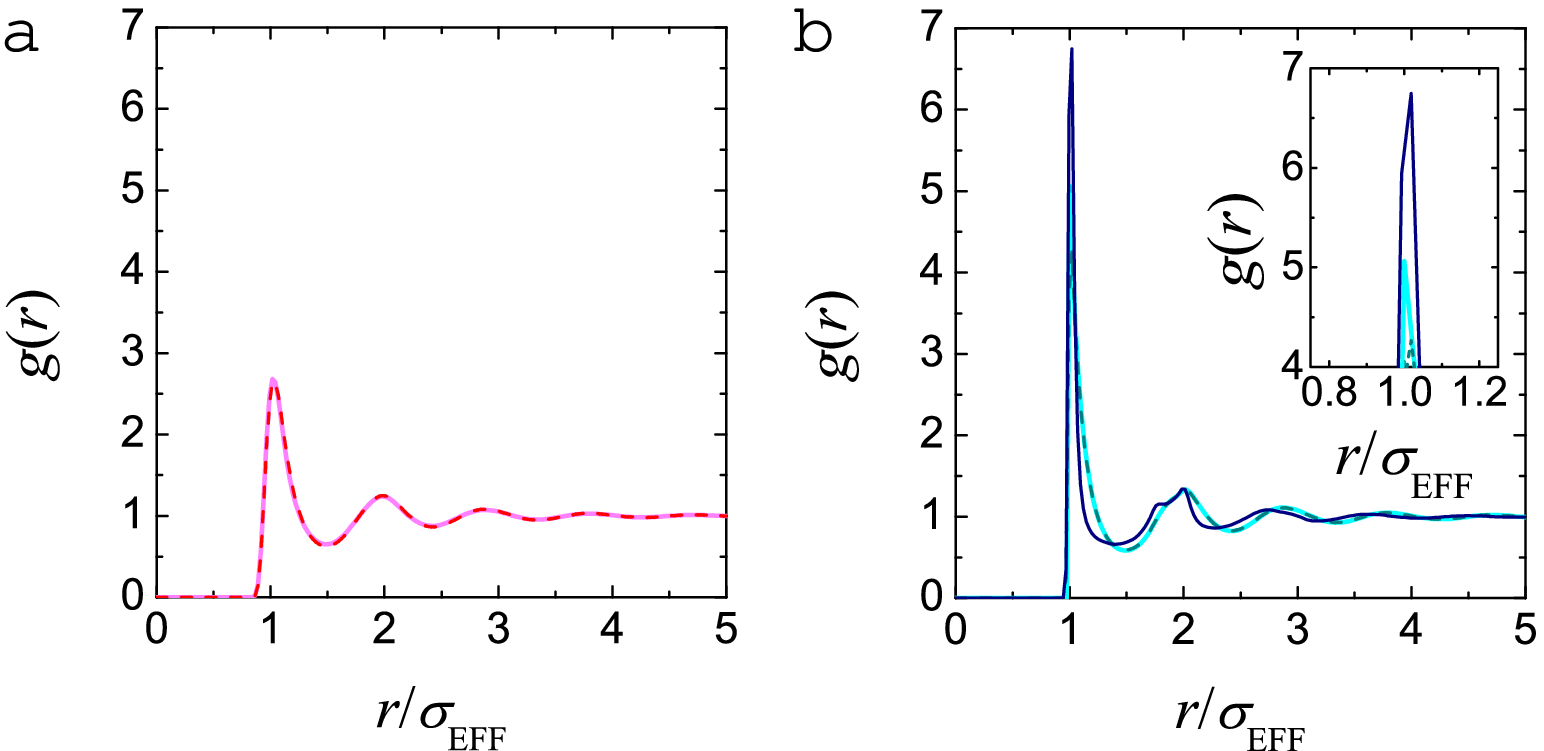} 
\par\end{centering}
\caption{(color online) Pair-correlation functions. (a) Long-ranged potentials:
Lennard-Jones (red) and WCA (pink) for a well depth of $\beta\varepsilon_{LJ}=0.4447$. 
(b) Short-ranged potentials: Morse (blue) and repulsive Morse (turquoise)
according to equation \ref{eqTMR}. Here the well depth $\beta\varepsilon_{M}=2.0$.
Cyan denotes the Hard Sphere interaction.}
\label{figG} 
\end{figure}

\begin{figure}
\begin{centering}
\includegraphics[width=15cm]{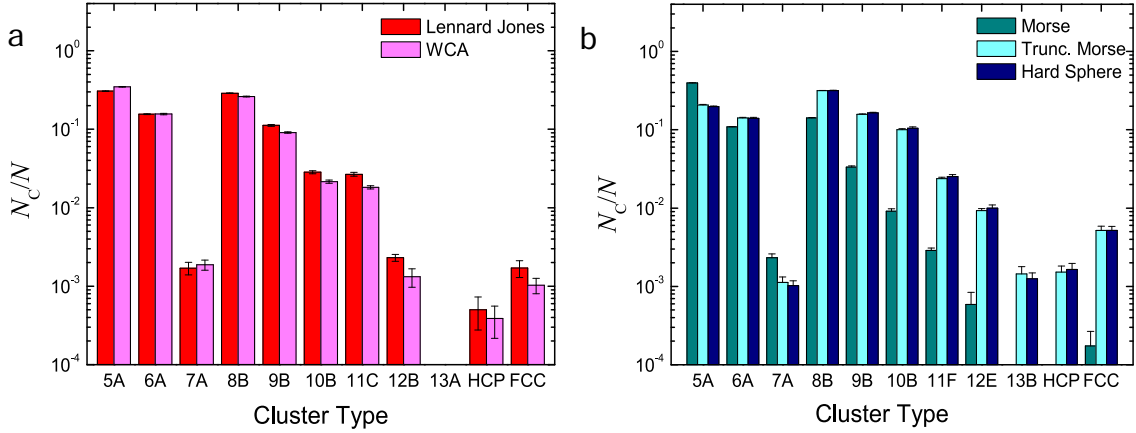} 
\par\end{centering}
\caption{(color online) Population of particles in a given ground state cluster. $N_c$ is the number of particles in a given cluster,
$N$ the total number of particles sampled. (a) Lennard-Jones with $\beta\varepsilon_{LJ}=0.440$ (red) and corresponding WCA (pink). (b) Morse ($\rho_0=25.0$) (turquoise)
truncated Morse (light blue) and hard sphere (dark blue). Note the semi-log scale.}
\label{figClusGround} 
\end{figure}

For shorter-ranged interactions relevant to colloid-polymer mixtures \cite{royall2008gel},
to avoid crystallisation we used
an attractive well depth of $\beta\varepsilon_M=2.0$, which corresponds via equation (\ref{eqCorrespondingStates}) to 
a Lennard-Jones well depth of $\beta\varepsilon_{LJ}\approx0.4447$. 
Pair correlation functions are shown in Fig. \ref{figG}. Again,
the WCA and Lennard-Jones [Fig. \ref{figG}(a)] show a similar behaviour.
In the case of the shorter-ranged potentials [Fig. \ref{figG}(b) and inset],
we see a strong increase in the first peak. The short-ranged Morse $\rho_0=25.0$ 
system shows some splitting of the second peak. We carefully
checked that no crystallisation was found during these simulation
runs, which is supported by the TCC analysis which shows a much \emph{reduced} 
population of particles in a locally crystalline environment upon adding attractions
(Fig. 
 \ref{figClusGround}). However, we are unaware of an equilibrium phase diagram for the Morse $\rho_0=25.0$
system, so we cannot exclude the possibility that the system is metastable to crystallisation, a point to which we return below. 
Furthermore, the first peak in $g(r)$ is rather higher in the case of the Morse interaction,
compared to the purely repulsive truncated Morse and hard-sphere interactions.
We note also that there is little difference between the truncated Morse and hard sphere pair correlation functions.
This suggests that the truncated Morse $\rho_0=25.0$ may provide a useful \emph{continuous} approximation to the hard sphere system.
We remark that the idea of the pair structure being dominated
by the hard core
\cite{weeks1971} appears less satisfactory here.

\begin{figure}
\begin{centering}
\includegraphics[width=15.0cm]{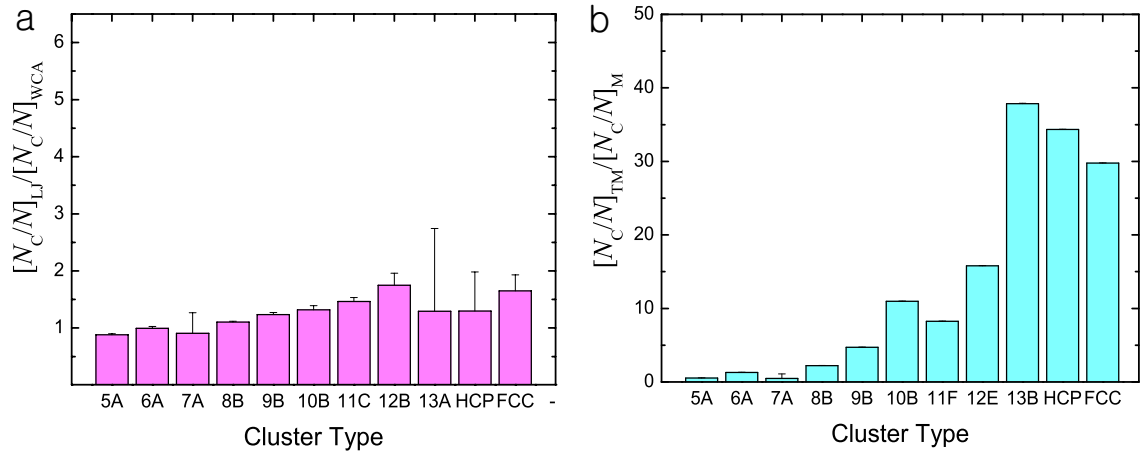} 
\par\end{centering}
\caption{(color online) Ratio of cluster populations in high temperature systems. (a) Lennard-Jones and WCA. (b) Truncated Morse and Morse. This plot is the same data as Fig. \ref{figClusGround} expressed to emphasize the difference between the two systems. Note that in (b) we invert the ratio to consider the truncated Morse divided by the attractive Morse potential and plot on a different scale.}
\label{figRatioShort} 
\end{figure}

We now turn our attention to the cluster populations in the Lennard-Jones-WCA
systems [Fig. \ref{figClusGround}(a)]. 
As before, we consider clusters
that are ground states for Lennard-Jones. 
At this higher temperature, compared to Fig. \ref{figClusGroundLong}(b),
relatively little difference
is seen between WCA and Lennard-Jones, consistent with the
concept that in dense  
liquids, it is the repulsions that are
responsible for the structure and that attractive interactions have less effect at higher temperature. 
However, the same trend is apparent as was found at the triple point [Fig. \ref{figClusGroundLong}(b)]:
Lennard-Jones shows a tendency to form larger clusters than
WCA, which seems reasonable given its cohesive energy and that
these clusters minimise the energy of isolated systems.
However, as shown by the ratio $[N_c/N]_{LJ}/[N_c/N]_{WCA}$ in Fig. \ref{figRatioShort}(a), the difference 
in population is slight.


For the truncated Morse and hard spheres,
the cluster populations are not tremendously different to the Lennard-Jones case, 
in fact for smaller clusters the differences are comparable to those between WCA and Lennard-Jones.
In particular, hard spheres show a very similar population to the truncated Morse, further suggesting that the 
latter might make a reasonable approximation to hard spheres.
However, upon adding attractions, the population of clusters drops dramatically. In Fig. \ref{figRatioShort}(b)
we plot the ratio $[N_c/N]_{TM}/[N_c/N]_{M}$ which is inverted with respect to Figs. \ref{figRatioShort}(a) and \ref{figRatio} in the sense that the truncated system forms the numerator and the attractive system forms the denominator. In the truncated Morse system, the population of clusters of size $m \geq 10$, is at least eight times
greater than the attractive system.
We return to the possible origins of this discrepancy in the next section. We note that the attractive Morse $g(r)$ exhibits a split second peak which the truncated Morse and hard spheres $g(r)$ do not. Yet, contrary to the notion that 
the split second peak is associated with crystallisation\cite{truskett1998}, in Fig. \ref{figClusGround}(b) we see precisely the opposite trend: that the split second peak is apparently associated with \emph{less} crystallinity. 

We now plot the population of all identified clusters, noting that it is only for $m\geq11$
that there is a difference in the ground state clusters for these interaction
ranges. That is to say, for Lennard-Jones, the ground states are 11C ($C_{2v}$),
12B ($C_{5v}$) and 13A icosahedron whereas for the short-ranged Morse the ground states are 11F ($C_{2v}$),
12E ($D_{3h}$) and 13B ($D_{5h}$). 
Excluding the attractive Morse system there is rather little variation between the different systems. In other words, it appears that for the other higher-temperature systems, the topological bond structure may be dominated
by the hard core, which is matched in all cases. In particular, although the Lennard-Jones and Morse potentials have different ground states, at these high temperatures this is little reflected in the structure.

%





\begin{figure}
\begin{centering}
\includegraphics[width=15cm]{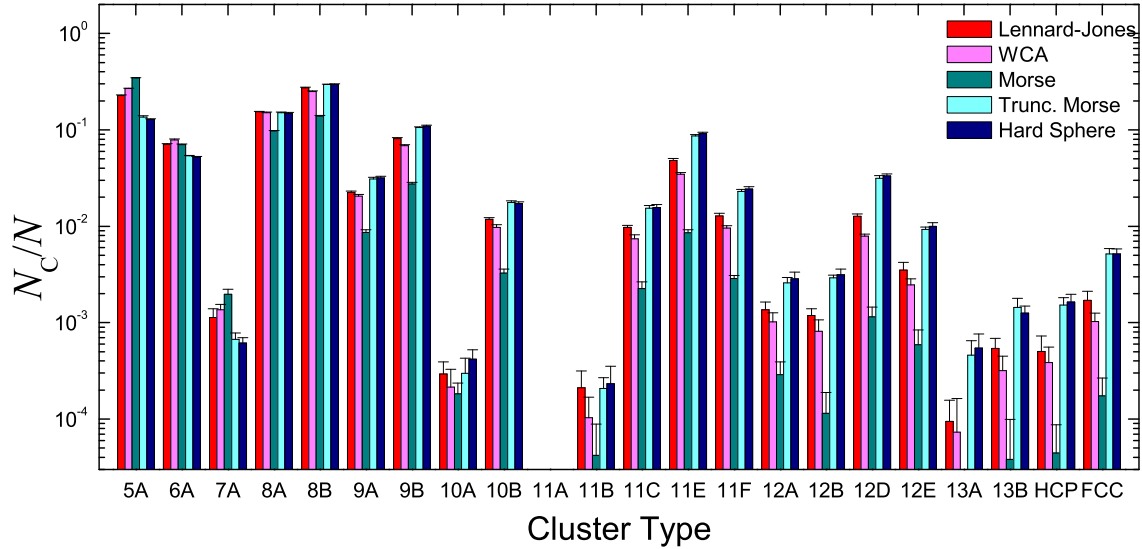} 
\par\end{centering}
\caption{(color online) Population of particles in a given cluster, at parameters mapped to the Morse potential ($\beta\epsilon_M=2.0$, $\rho_0=25.0$). $N_c$ is the number of particles in a given cluster,
$N$ the total number of particles sampled. Here we consider ground state clusters for all ranges of the Morse potential \cite{doye1995}. 
Note the semi-log scale.}
\label{figAllShortHotSort} 
\end{figure}

\section{Discussion and Conclusions}

\label{sec:Conclusions}

We have analysed the pair structure and performed a topological cluster classification on a range of liquids. 
The pair structure of Lennard-Jones and longer-ranged liquids is entirely 
consistent with the well-known result that repulsive interactions dominate the local packing in dense liquids.
Shorter-ranged potentials exhibit a strong response in the $g(r)$ upon the addition of attractions. However, 
one expects that these will be accounted
 for by the use of perturbation theory \cite{weeks1971}.
We note that as $\rho \rightarrow 0$, $g(r) \rightarrow \exp[-\beta u(r)]$ so a short-ranged
attraction leads to a strong peak at contact as we see in Figure \ref{figG}(b).

Although the pair structure of Lennard-Jones and WCA is very similar, we are nonetheless able to identify clear differences
using the TCC. We find that Lennard-Jones is more able to form higher-order clusters than the purely repulsive WCA. These
differences become much more significant upon cooling to the triple point. Applying the extended law of corresponding states to compare with a weakly interacting longer-ranged system, little effect on the cluster population, or pair correlation function 
is found upon adding attractions.
Conversely, in short-ranged systems, the radial distribution function is influenced by attractions 
and the cluster population is strongly enhanced upon \emph{removing}
attraction. That we see different trends in the short-ranged system is rather curious, and will be investigated further in the future. One comment we can make at this stage is that short-ranged attractive systems exhibit non-monotonic dynamics as a function of attraction at high densities, in the form of a re-entrant glass transition \cite{pham2002, zaccarelli2002, krekelberg2007}. Whether it is truly appropriate to expect the behaviour of short-ranged attractive systems to be similar to long-ranged Lennard-Jones type liquids is perhaps an open question. 

We rationalise these three scenarios as follows. The long-ranged Morse is weakly interacting and compressed. Together, these lead to little response either of the $g(r)$ (the spatial distribution of particles) or the topology upon adding attractions. In the Lennard-Jones case compression is less important, adding attraction promotes organisation and clustering, however the interactions are sufficiently long-ranged that the repulsive core dominates the pair structure for both Lennard-Jones and WCA. In the short-ranged case, the hard spheres (and presumably the truncated Morse) are close to freezing (here the packing fraction $\phi=0.4875$) and thus have limited free volume. 
Adding short-ranged attraction favours configurations where the particles are closer to contact,  raising the first peak of $g(r)$, and can 
open up free volume. However, considering the second Virial coefficents $B^*_2$ in table \ref{tableStatePoints}, these short-ranged systems are quite weakly interacting, and there is apparently insufficient cohesive energy to promote organisation into clusters.

Returning to the non-monotonic dynamics of short-ranged systems, one expects that 
the attractive Morse system might exhibit faster dynamics than hard spheres (and perhaps the truncated Morse system). Although the hard sphere packing is far from dynamical arrest, even so some kind of slowing is expected relative to a dilute fluid. This could then be reduced by the short-ranged attraction. Now we have correlated the clusters with dynamical arrest 
\cite{royall2008gel} and found that arrested states have a high cluster population and that it is biased towards higher-order clusters. As a function of density, hard sphere fluids show a similar trend \cite{williams2007}. Thus we speculate that one possible underlying cause may be related to dynamics. However, the hard sphere packing fraction is very close to freezing, and we note a substantial quantity of locally crystalline particles in Fig. \ref{figClusGround}(b). Now the colloid-polymer literature \cite{poon2002} would tend to suggest that adding short-ranged
attractions widens the fluid-crystal coexistence region. Since the hard sphere state point is so close to freezing, and the truncated Morse seems similar to hard spheres, it is possible that the equilibrium state for Morse ($\rho_0=25.0, \beta \varepsilon_M=2.0$) is crystal-fluid coexistence. It is interesting to note that this possibly metastable fluid has a population of HCP and FCC structures around a factor of $30$ less than the stable hard sphere fluid. 


Among the key underlying ideas of clusters in liquids is that they represent energetically locally favoured structures \cite{frank1952, coslovich2007, mossa2003}. The most famous of these, the icosahedron, appears only in small quantities in this analysis, although it is most prevalent in Lennard-Jones.
It would be most interesting to investigate whether particles in these clusters are in fact in a low energy environment. It would also be interesting to seek a link between structure and dynamics, particularly concerning the recent observation of very different dynamical behaviour between the WCA and Lennard-Jones systems
\cite{berthier2009} and the observation that power-law repulsive interactions seem to recapture the original Lennard-Jones behaviour \cite{pedersen2010}. Moreover, other mappings have been proposed for example between Lennard-Jones
and WCA. Here one can place more emphasis upon the dynamics, albeit at some expense in the accuracy with which the 
radial distribution function is matched \cite{young2005}.

Here we have focused on the ground state clusters for each system. Furthermore, liquids are by definition at finite temperature, and it may be appropriate to consider the structure of clusters at higher temperature in addition to the ground states we have investigated so far \cite{honeycutt1987, baletto2005, berry1994, malins2009}. Conversely, further quenching might favour the ground states beyond the trends we have so far seen. Recently, we found we needed around $10k_BT$ of attraction to form isolated clusters \cite{malins2009}.

A final point for discussion is the link between attractions and reciprocal space structure. The static structure factor $S(q)$ measures compressibility
at wavevector $q=0$. While all state points sampled show no indication of sny phase transition, one nonetheless expects some hint of attractions at low $q$ \cite{evans1981, stell1980}. In particular, around the Lennard-Jones triple point, 
for $q \rightarrow 0$ we might expect a factor of two increase in $S(q)$ between WCA and LJ \cite{stell1980}. This difference in $S(q)$ upon
adding attractions is predicted to be most prevalent at a value of $q \sigma \approx 1$ or around $2 \pi \sigma$ in real space. This is a rather larger lengthscale than the clusters we probe, and in fact Stell and Weis \cite{stell1980} show that by $q \sigma \approx 3$, at the length scales we consider here, this effect is much reduced. It would be most interesting to extend the TCC to larger clusters, such that the $q\sigma \approx 1$ range might be reached. Thus we could directly investigate the nature of the change in structure related to this low $q$ behaviour. However, we note that the TCC is a particle-based methodology, and may not be sensitive to such delicate changes in long-ranged structure.

\section*{Acknowledgements}

JT and CPR thank the Royal Society for funding, AM is supported under
EPSRC grant EP/E501214/1. The authors are grateful to Jens Eggers, Bob Evans,
Rob Jack and an anonymous reviewer for helpful discussions and suggestions.

\newpage

\section*{References}

\bibliography{jade}

\newpage

\end{document}